\documentclass[fleqn,usenatbib]{mnras}

\usepackage{newtxtext,newtxmath}

\usepackage[T1]{fontenc}

\DeclareRobustCommand{\VAN}[3]{#2}
\let\VANthebibliography\thebibliography
\def\thebibliography{\DeclareRobustCommand{\VAN}[3]{##3}\VANthebibliography}

\usepackage{ae,aecompl}
\usepackage{soul}
\soulregister\cite7 
\soulregister\citep7 
\soulregister\citet7 
\soulregister\ref7 
\soulregister\pageref7 

\usepackage{graphicx}	
\usepackage{amsmath}	
\usepackage{amssymb}	
\usepackage{arydshln}

\title[Using Gaia DR2 to solve DCR and CTE issues]{
Using Gaia DR2 to solve differential color refraction and charge transfer efficiency issues}

\author[F. R. Lin et al.]{
F. R. Lin,$^{1,2}$
Q. Y. Peng$^{1,2}$\thanks{tpengqy@jnu.edu.cn} and
Z. J. Zheng$^{1,2}$
\\
$^{1}$Department of Computer Science, Jinan University, Guangzhou 510632, China\\
$^{2}$Sino-French Joint Laboratory for Astrometry, Dynamics and Space Science, Jinan University, Guangzhou 510632, China
}

\pubyear{2020}

\begin{document}
\label{firstpage}
\pagerange{\pageref{firstpage}--\pageref{lastpage}}
\maketitle

\begin{abstract}
The Gaia DR2 catalog released in 2018 gives information about more than one billion stars, including their extremely precise positions that are not affected by the atmosphere, as well as the magnitudes in the $G$, $RP$, and $BP$ passbands. This information provides great potential for the improvement of the ground-based astrometry. Based on Gaia DR2, we present a convenient method to calibrate the differential color refraction (DCR). This method only requires observations with dozens of stars taken through a selected filter.  Applying this method to the reduction of observations captured through different filters by the 1-m and 2.4-m telescopes at Yunnan Observatory, the results show that the mean of the residuals between observed and computed positions $(O-C)$ after DCR correction is significantly reduced. For our observations taken through an $N$ (null) filter, the median of the mean $(O-C)$ for well-exposed stars (about 15 $G$-mag) decreases from 19 mas to 3 mas, thus achieving better accuracy, i.e. mean $(O-C)$.
Another issue correlated is a systematic error caused by the poor charge transfer efficiency (CTE) when a CCD frame is read out. This systematic error is significant for some of the observations captured by the 1-m telescope at Yunnan Observatory. Using a sigmoidal function to fit and correct the mean $(O-C)$, a systematic error up to 30 mas can be eliminated.
\end{abstract}

\begin{keywords}
astrometry - techniques: image processing - methods: statistical
\end{keywords}


\section{Introduction}           
\label{sect:intro}
With the improvement of the orbital theory of natural satellites, even slight systematic errors would be detected in the astrometric results, which puts forward stricter requirements for astrometric techniques. 
Moreover, using the newly released high-precision catalog Gaia DR2 \citep{brown2018gaia}, the ground-based astrometry has been greatly improved in both precision and accuracy. These developments highlight the importance of some previously negligible issues, the effect of differential color refraction (DCR) is one of them.

It is well known that the refractive index of the atmosphere depends on the wavelength. The light of shorter wavelength is more refracted and the direction of the refraction is toward the zenith. As a result, the light from a star will be refracted into a spectrum when passing through the Earth's atmosphere. In other words, stars with different spectral types will experience differing degrees of atmospheric refraction, which is referred to as differential color refraction \citep{stone2002new}.
Differential color refraction may result in a systematic error in the zenith direction, so it should always be considered in ground-based astrometry in order to obtain the high-accuracy position of a star \citep{anderson2006ground, velasco2016high}. \citet{monet1992us} proposed a method to calibrate DCR when they measured the parallax of stars. This method requires to observe the same field of view (FOV) at different times in a single night, and obtain a series of observations when the target FOV is located at different zenith distances (ZD) to determine the DCR effect. Then, the DCR can be calibrated after obtaining the color index of stars by the photometry of them. Using this method to process the observations of 166 stars captured in a span of about four hours, the slopes (relationship between the DCR effects and star colors) for individual stars were measured with an uncertainty of about $\pm10\%$ \citep{monet1992us}. The method is commonly used and has successfully computed DCR in some works \citep[e.g.][to name a few]{tinney1993faintest,ducourant2008accurate,velasco2016high}. Another method to calibrate DCR was presented by \citet{stone2002new}. Stone used the H$\alpha$ interference filter, which has a very narrow passband, to obtain observations that can be treated as monochromatic (its DCR is only 50 $\mu$as). At the same time, observations of the same FOV were captured in turn through Johnson $BV$, Cousins $RI$ filters and the H$\alpha$ filter in a successive period of time. The pixel coordinates of observations in Johnson $BV$ and Cousins $RI$ passbands were mapped onto the H$\alpha$ positions separately, then the DCR in each passband can be calculated according to the relationship between the color index of each star and its mapping residual. Obviously, both methods require some extra observations to determine the color index of stars and DCR effects. Sometimes, however, the limitation of observation time, instruments or weather conditions makes it impossible to apply the methods mentioned above. Many works have to choose to minimize the effect of DCR by observing a target at a small ZD and using an appropriate filter (such as in $I$ or $K$ band), so that high-accuracy results can be obtained even though the DCR is not corrected \citep[e.g.][]{neuhauser2008astrometric,kilic201211,dieterich2018dynamical,wang2018focus}.

Unfortunately, the observations of some targets will inevitably be affected by DCR. For example, when observing the Galilean satellites of Jupiter, Johnson $B$ filter would usually be selected to alleviate the image oversaturation caused by the high luminosity of Jupiter \citep{peng2003image,peng2012precise}. At the same time, the target may have a large ZD in the observable period. For example, from 2018 to 2021, the ZD of a Jupiter's satellite is larger than 40$^\circ$ when observing at Yunnan Observatory, so its positional measurement would be affected by DCR obviously, and the systematic error can be as large as dozens of mas.
In the past, the astrometric accuracy of moving targets was not very high due to various reasons, such as the irregularity of the figure of the satellites \citep{hog2014solar}, the error of the catalog or ephemeris, etc. Therefore, the DCR effect was usually not considered in the observation of satellites and asteroids \citep[e.g.][]{gomes2015astrometric,wang2015precise,yu2018new}.
But now the release of the Gaia star catalog has caused a revolution in astrometry. It provides very precise positions (including parallax, proper motion and radial velocity and so on), as well as high-precision photometry in the $G$, $BP$ and $RP$ passbands. With the high-precision Gaia catalog, both the measurement precision and accuracy of these targets can be obviously improved \citep{peng2017precise,wang2017precise}, and thus more stringent requirements are put forward for astrometric techniques.

 In this paper, we investigate a convenient method to calibrate DCR based on the astrometry and photometry provided by Gaia DR2. This method only requires the observations with dozens of stars to calibrate DCR of the filter used, thus the situations mentioned above can be handled well.
The practicability of this method for different filters is tested using observations taken through the $N$ (null) and $BVRI$ filters by the 1-m and 2.4-m telescopes at Yunnan Observatory. Specifically, we calibrate the DCR of each filter against color index $BP-RP$, and the correction for DCR is applied in the reduction of these observations. Then the DCR corrected results are compared with the results of observations taken under similar observational conditions (i.e. seeing, temperature, etc.), but through $I$ filter at small ZD. Besides, the mean $(O-C)$ (denoted as $\langle O-C\rangle$ hereafter) of DCR corrected results in altitude and azimuth are compared to demonstrate that the effect of DCR can be effectively eliminated.

The use of Gaia catalog makes it much easier to solve another issue, namely the systematic error caused by poor charge transfer efficiency (CTE). At the readout stage of an exposure, a few electrons are left behind as the charge packet moves from pixel to pixel, which results in the CTE issue. This issue has existed since CCD was used as the detector, it leads to the $\langle O-C\rangle$ systematically changing with the magnitude of stars \citep{hoist1996ccd}. In general, a plate constant model with magnitude terms could be used to handle this error \citep{finch2010ucac3,robert2011new,robert2014astrometric}, or polynomial could be used to fit the $\langle O-C\rangle$s and remove the systematic error from them \citep{zacharias2000first}. With the improvement of CCD performance, this error is gradually submerged in the error of the catalog and so cannot be detected in our reduction before. However, when Gaia DR2 is used as the reference star catalog, we find that the systematic error associated with the magnitude appears in the reduction of some observations captured by the 1-m telescope at Yunnan Observatory. That is to say, the effect of CTE can now be accurately determined. And the CTE issue should be taken into account to obtain the improved results. In this paper, the methods to deal with the CTE issue are investigated using observations of different targets.

The contents of this paper are arranged as follows. In Section 2, the observations used to compute DCR and study the CTE issue are presented in detail. The methods to eliminate errors caused by DCR and CTE are given in Section 3. Section 4 shows the results after DCR and CTE corrections, the detailed parameters of DCR solutions for different filters of the 1-m and 2.4-m telescopes at Yunnan Observatory are also given in this section. Finally, some conclusions are drawn in Section 5.
\section{Observations}
\label{sect:obs}
In this paper, DCR calibrations are carried out for observations taken through different filters by the 1-m telescope at Yunnan Observatory (IAU code 286, longitude\---E$102^{\circ}47^{\prime}18^{\prime\prime}$, latitude--N$25^{\circ}1^{\prime}30^{\prime\prime}$, and height--2000m above sea level). Besides, four-night observations captured by the 2.4-m telescope (IAU code O44, longitude\---E$100^{\circ}1^{\prime}51^{\prime\prime}$, latitude--N$26^{\circ}42^{\prime}32^{\prime\prime}$, and height--3193m above sea level) at Yunnan Observatory through its $B$ filter are used to test the stability of the DCR solution. Details of these observations are given in Table~\ref{tbl1}. Among them, the observations of open clusters (NGC2324, M35, M39, M67, NGC6633 and NGC6709) and a dense star field around the star HIP91882 were originally captured for geometric distortion correction. They were taken by the dithered observational scheme (``+'' or ``\#'' type), which were presented in \citet{peng2012convenient}. The target Apophis is a fast-moving near Earth asteroid, so the observational FOV would also change significantly with the movement of the target. Therefore, these observations contain more reference stars. The observations of target AH Aur were originally captured for astrophysics, they were taken through different filters and used in this paper for DCR calibration. These observations have fixed FOV and therefore have fewer reference stars to compute DCR. Nevertheless, the impact of DCR to all of these observations can be well eliminated by our method (see Section~\ref{sect:med}). More instrumental details of the reflectors and CCD detectors are listed in Table~\ref{tbl2}. In addition, the information of two observation sets obviously affected by CTE issue is given in the last two lines of the Table~\ref{tbl1}.

\begin{table*}
\begin{center}
	\caption{Details of the observations captured by the 1-m and 2.4-m telescopes at Yunnan Observatory.\label{tbl1}. The first column is the identification of the observation sets. Column (2) and Column (3) list the target and the zenith distance of the observations respectively. Column (4) is the observational date. The observations could be used together to compute DCR if they were captured through the same filter in a short period of time. Column (5) gives the number of CCD frames in each observation set. Column (6) is the filter used to capture the observations. Column (7) shows which telescope is used and column (8) gives the total number of stars observed in each observation set. Column (9) gives the exposure time.
}
\begin{tabular}{@{}c*{15}{c}}
\hline\hline
ID&Targets&ZD (mean)&Obs Date&No.&Filter&Telescope&Stars&Exposure \\
  &             &(deg)    &(y-m-d) &   &      &         &     & (second)\\
 (1)&(2)&(3)&(4)&(5)&(6)&(7)&(8)&(9)  \\
\hline
1  & NGC2324    & 25-28 (25.3)& 2011-02-26    & 58    & $N$     & 1-m   & 1656&60 \\
2  & AH Aur     & 20-41 (30.7)& 2012-10-14    & 357   & $N$     & 1-m   & 116 &10 \\
3  & HIP91882   & 14-28 (19.7)& 2013-05-13    & 93    & $N$     & 1-m   & 862 &20 \\
4  & M39        & 24-34 (26.9)& 2014-10-17,18 & 37    & $N$     & 1-m   & 882 &60 \\
5  & M35        & 2-22 (11.2)& 2017-11-12    & 44    & $I$     & 1-m   & 1102 &50   \\
6  & NGC6709    & 18-52 (34.8)& 2012-04-18    & 86    & $I$     & 1-m   &760  &60 \\
7  & NGC6633    & 29-32 (30.0)& 2017-05-28,29 & 23    & $R$     & 1-m   &509  &60 \\
8  & AH Aur     & 3-37 (19.2)& 2015-12-21    & 157    & $V$     & 1-m   &89   &10 \\
9  & NGC2324    & 30-42 (35.5)& 2018-03-22    & 26    & $B$     & 1-m   &195  &100 \\
10 & M67        & 14-26 (17.0)& 2019-03-30,31    & 21    & $V$     & 1-m   &156  &60 \\
11 & Apophis    & 36-33 (35.0)& 2013-02-04    & 36    & $B$     & 2.4-m &681  &25 \\
12 & Apophis    & 39-33 (34.0)& 2013-02-05    & 56    & $B$     & 2.4-m &1756 &30 \\
13 & Apophis    & 51-32 (36.4)& 2013-02-06    & 64    & $B$     & 2.4-m &1586 &30 \\
14 & Apophis    & 36-32 (33.7)& 2013-02-07    & 30    & $B$     & 2.4-m &674  &30 \\
\cdashline{1-9}[3pt/4 pt]
15 & AH Aur     & 5-37  (20.4)& 2017-11-14    & 175    & $I$     & 1-m &114  &5\\
16 & M35        & 10-36 (23.7)& 2019-11-23    & 49    & $I$     & 1-m &1228   &60\\
\hline
\end{tabular}
\end{center}
\end{table*}

\begin{table*}
\begin{center}
\caption{Specifications of the 1-m and 2.4-m telescopes and the corresponding CCD detectors.\label{tbl2}}
\begin{tabular}{@{}l*{15}{l}}
\hline\hline
Parameter&1-m telescope&2.4-m telescope\\
\hline
Approximate focal length&1330 cm &1920 cm \\
F-Ratio & 13 &8 \\
Diameter of primary mirror&100 cm & 240 cm \\
Approximate scale factor & 0.209 arcsec pixel$^{-1}$ & 0.286 arcsec pixel$^{-1}$\\
Size of CCD array (effective)& 2048 $\times$ 2048&1900 $\times$ 1900 (cropped) \\
Size of pixel &13.5 $\mu m\ \times$ 13.5 $\mu m$ & 13.5 $ \mu m\ \times$ 13.5 $\mu m$\\
\hline
\end{tabular}
\end{center}
\end{table*}

\section{Methods}
\label{sect:med}
The procedures to compute DCR are outlined as follows. The calibration stars in the observations are firstly matched to the Gaia DR2 catalog to obtain their astrometric and photometric data. Then the astrometric places in the equatorial coordinates are transformed to the topocentric apparent places (including atmosphere refraction) in the alt-azimuth coordinates \citep{green1985spherical,kaplan1989mean}. So we have the altitude and azimuth of each star. Then, a weighted least squares scheme described in \citet{lin2019characterization} is used to solve the plate model of the alt-azimuth system. Specifically, the weight of each star in the least squares solution is designed by
\begin{equation}
\label{eq1}
w_p=1/\sigma^2(m),
\end{equation}
where $m$ is the magnitude of the star and $\sigma(m)$ is a function describing the relation between its magnitude and its measurement precision. $\sigma(m)$ can be expressed as a sigmoidal function
\begin{equation}
\label{eq2}
\sigma(m)=(A_1-A_2)/(1+e^{(m-m_0)/dm})+A_2,
\end{equation}
where $A_1$ and $A_2$ represent the initial and final values of the sigmoidal curve respectively, $m_0$ is the $m$ value of the curve's midpoint and $dm$ the logistic growth rate or steepness of the curve \citep{verhulst1838notice}. The curves fitted by Equation~\ref{eq2} are plotted in Figure~\ref{fig3} and Figure~\ref{fig5}, and the corresponding fitting parameters are given in their captions. The sigmoidal curve is used here since it was found to be suitable for describing the astrometric precision in our previous work, and residuals of the curve fitting were also given in that work to show the quality of the fitting \citep[see Figure 1 in][]{lin2019characterization}.

The order of the plate model is selected according to the number of stars in each CCD frame. In this paper, the plate model up to fourth-order is used in the reduction of the 1-m telescope observations when there are enough reference stars according to our previous experiment\citep{peng2010accurate}. For the observations of the 2.4-m telescope, the plate model of no less than third-order is adopted to avoid the effects of geometric distortion. Otherwise, the geometric distortion should be corrected first by the method given in \citet{peng2012convenient}.

Now the $(O-C)$ residuals in alt-azimuth coordinates can be computed using some plate model. If the observation lasts for a period of time in a night, the effect of DCR may change with frames. At this time, the $(O-C)$ residuals should be normalized according to the DCR effect on them to derive a more precise DCR solution. This can be done by using the equation \citep{stone2002new}
 \begin{equation}
\label{eq3}
\operatorname{DCR}\left(t^{\prime}, P^{\prime}, \mathrm{ZD}^{\prime}\right)=\frac{t+273.15}{t^{\prime}+273.15} \frac{P^{\prime}}{P} \frac{\tan \mathrm{ZD}^{\prime}}{\tan \mathrm{ZD}} \mathrm{DCR}(t, P, \mathrm{ZD}),
\end{equation}
which converts the effect of DCR with the centigrade temperature $t$, pressure $P$ and zenith distance ZD to that with $t^{\prime}$, $P^{\prime}$ and ZD$^{\prime}$. Parameters $t$ and $P$ are found not to change significantly in a single night, so they can be ignored here.
Selecting the appropriate color index $BP-RP$ given by Gaia DR2, the relationship between the $(O-C)$ residual and the star color can be determined by a polynomial fitting. Since the residual has been scaled according to the ZD, an additional weight should be used in the least squares fitting, which is computed by
\begin{equation}
\label{eq6}
w_s=\tan^2\mathrm{ZD}/\sigma^2(m).
\end{equation}
The factor $\tan^2(\mathrm{ZD})$ is introduced into the weight since the $(O-C)$ residuals were normalized according to Equation~\ref{eq3}. The first-order polynomial is found enough to fit the effect of DCR on all stars except few stars with extreme colors (less than 1$\%$ in our observations). Hence we can obtain a DCR solution expressed by
\begin{equation}
\label{eq5}
\operatorname{DCR}\left(a_{1}, a_{2}\right)=a_{1}+a_{2}\cdot color\cdot \tan \mathrm{ZD},
\end{equation}
where $color$ and ZD are the color index ($BP-RP$) and zenith distance of star respectively, and $a_{1}$ and $a_{2}$ are the fitted DCR parameters.
The effect of DCR on altitude is decomposed into the pixel coordinates (i.e. $x$ and $y$) of a CCD frame, and then the DCR-corrected pixel coordinates of stars can be calculated. We correct DCR in pixel coordinates instead of equatorial coordinates because the DCR effects are correlated with the instruments. Correction in pixel coordinates would demonstrate more clearly the physical meaning, which has also been considered in geometric distortion correction described in \citet{peng2012convenient}.

The systematic error caused by poor CTE is also related to the color of stars to some extent \citep{stone2002new}, so it may be confused with the solution of DCR. Fortunately, this error only occasionally appears in observations, it is not hard to select the observations which are not affected by this issue to compute DCR.
If the $\langle O-C\rangle$s of DCR corrected results are symmetrically distributed around 0 against the magnitudes, then we consider that the observations are not affected by CTE issue.
Moreover, some simple observational scheme can be used to ensure the DCR solution not affected by CTE issue, e.g. \citet{stone2002new} proposed to obtain the observations of different orientations by rotating the CCD 180$^\circ$, then the effect of CTE can be eliminated by average the $(O-C)$ residuals of two measurements taken at different CCD rotation.

Once the DCR of the observations has been well calibrated, it is relatively simple to deal with CTE issue only.
The plate model with magnitude terms can be used to deal with the CTE issue in observations of dense star fields. For the observations of sparse star fields, using magnitude terms in the plate model fitting will lead to over fitting problem. At this time, Equation~\ref{eq2} can be used to fit the $\langle O-C\rangle$s in the readout direction of the CCD, and the error caused by poor CTE can be eliminated by subtracting the fitting value from the $\langle O-C\rangle$. In addition, we found that the effect of CTE sometimes would change with time during the observation period of a single night, which is perhaps caused by the change of seeing or CCD operating temperature \citep{bautz2019toward}. Therefore, the plate model with magnitude terms is preferable to reach higher precision as long as there are enough well-exposed calibration stars (several hundred in a frame).

\section{Results}
\label{sect:res}
\subsection{Results of DCR correction}
\begin{figure}	
	\centering
\includegraphics[width=0.46\textwidth]{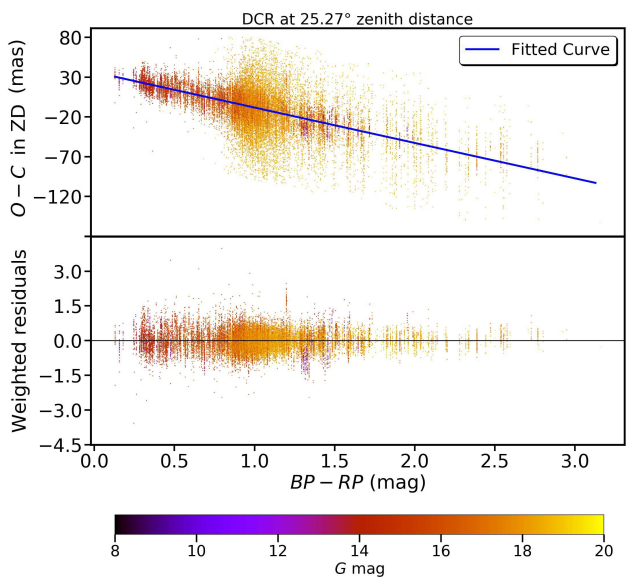}
		\caption{Upper panel: the $(O-C)$ residuals change with the color index $BP-RP$ for observation set 1, the blue line in the upper panel represents the fitted result of Equation~\ref{eq5}. The $(O-C)$ residuals are converted to 25.27$^\circ$ zenith distance (ZD) and outliers in the fitting process have been discarded using RANSAC \citep{fischler1981random} algorithm. Here the 25.27$^\circ$ is the mean ZD of the observation set. Lower panel: the residuals of the DCR function fitting. Since the weight expressed by Equation~\ref{eq6} is used in the least squares fitting, the vertical axis only represents a weighted residual and there's no unit for it. The color of the dot represents the Gaia $G$-mag of a star.} 	
	\label{fig1}
\end{figure}

Almost all light in the optical band is allowed when no filter is used during the observations, and so star images with higher signal-to-noise ratio (SNR) can be obtained. As a result, a null filter is usually used in the observation of targets with low to medium brightness, especially when the seeing is poor, so as to reduce centering error and improve measurement precision. Observation sets 1 to 4 given in Table~\ref{tbl1} are taken through the null filter. However, since the null filter has a very broad passband, these observations would be seriously affected by DCR even when they are taken at a zenith distance below 30$^\circ$.
Figure~\ref{fig1} shows the $(O-C)$ residuals change with the color index $BP-RP$ for observation set 1, and the weighted residuals of the DCR function fitting (lower panel). It is obvious that there is a systematic error of more than 100 mas in these $(O-C)$ residuals. To eliminate this error, we apply the DCR correction method described in Section~\ref{sect:med} to the observations. The parameters of DCR equation derived are listed in the first line of Table~\ref{tbl3}, parameters derived from other observations affected by DCR are also given in that table.

It should be noted that DCR only affects the altitude in the alt-azimuth system, and has no effect on the azimuth measurement. However, the right ascension and declination coordinates in the equatorial system are usually required in practical astrometric applications. The latter two coordinates usually be affected by DCR together, so DCR correction will change the measurement results in both directions. For convenience, we use the median of the statistics $\langle O-C\rangle_{sum}=\scriptstyle{\sqrt{\langle O-C\rangle^2_\alpha+\langle O-C\rangle^2_\delta}}$ for the comparison of change in the mean residual before and after systematic error correction. The median can reflect the dispersion of the mean residual and thus the accuracy of the measurement results.

\begin{figure*}	
	\centering
\includegraphics[width=0.49\textwidth]{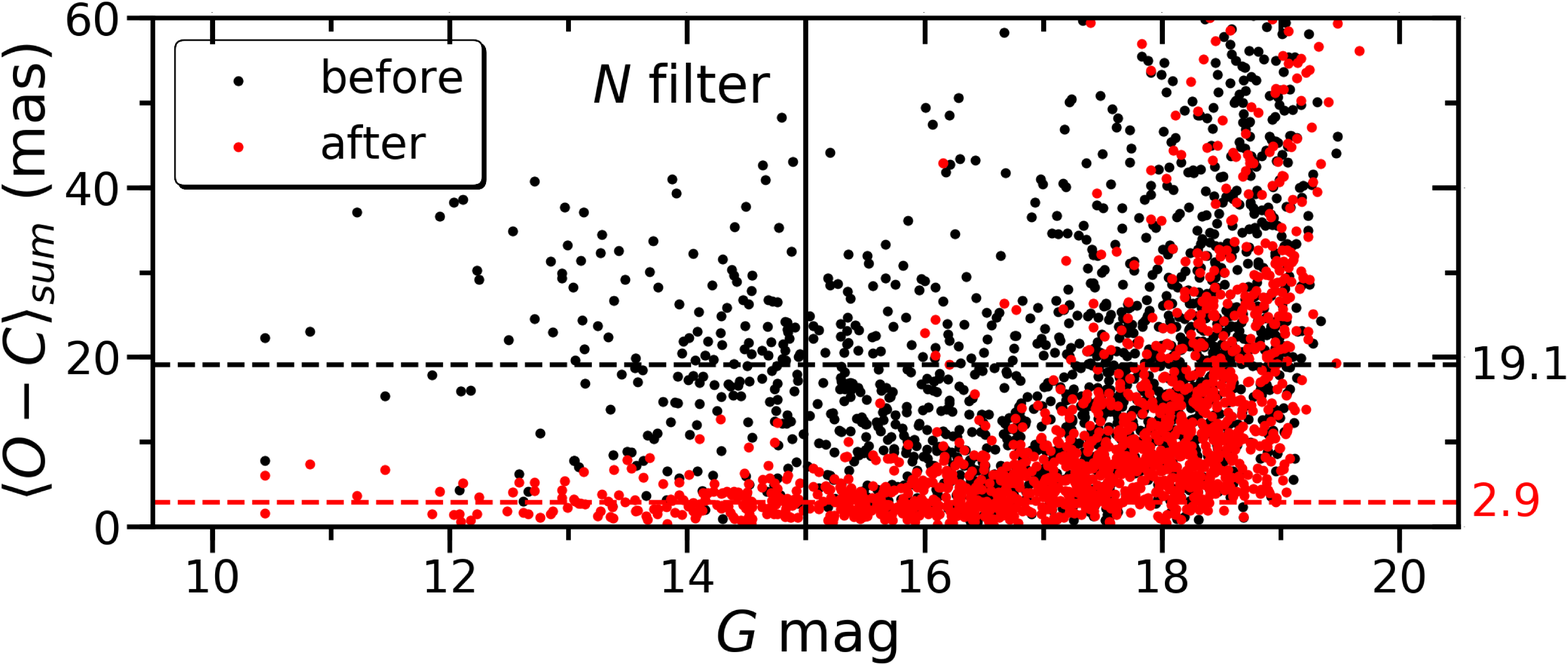}
\includegraphics[width=0.482\textwidth]{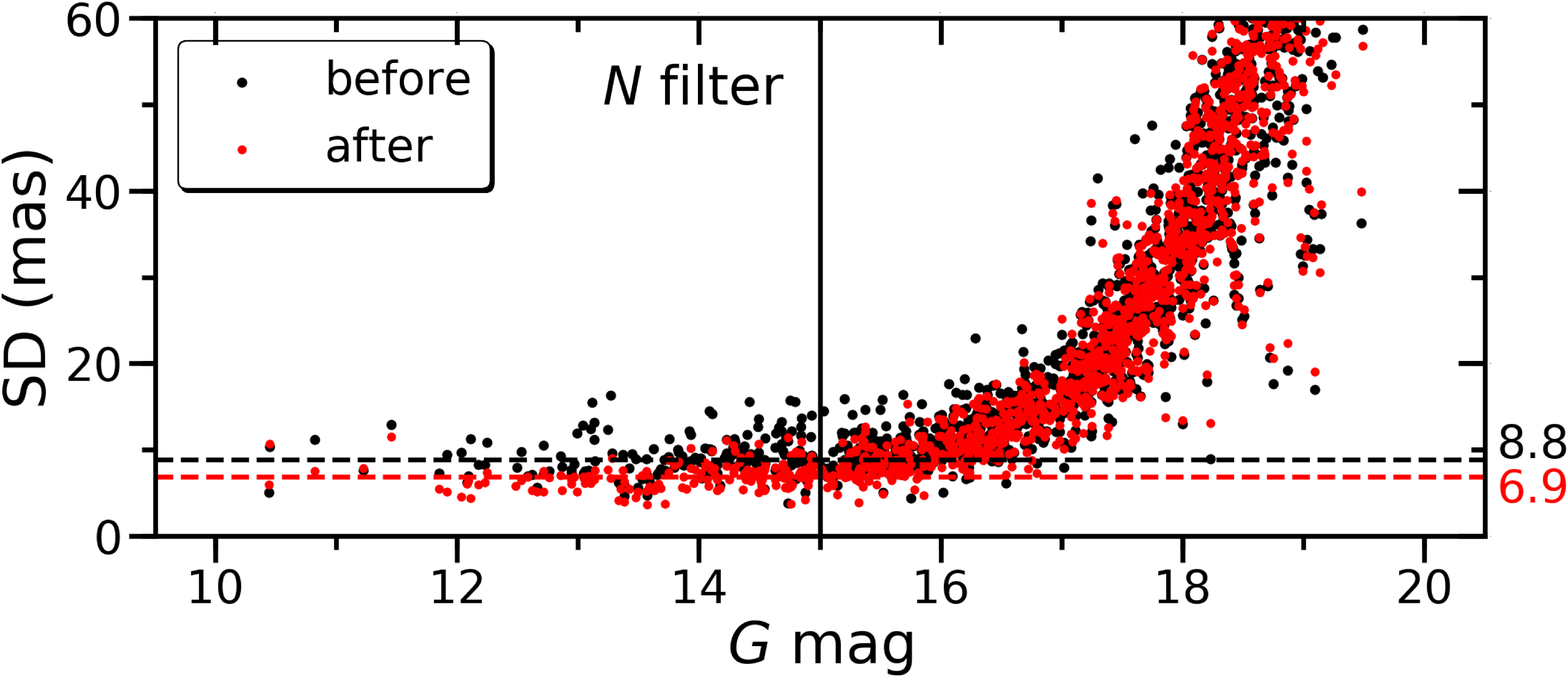}
		\caption{The statistics of the $(O-C)$s in the equatorial system before and after DCR correction for observation set 1. Left panel: the black dots and red dots are the statistics $\scriptstyle{\langle O-C\rangle_{sum}=\sqrt{\langle O-C\rangle^2_\alpha+\langle O-C\rangle^2_\delta}}$ before and after DCR correction respectively. The dash lines represent the medians of these statistics for stars brighter than 15 $G$-mag (marked by a vertical line). Right panel: positional standard deviation (SD), which is calculated by $\scriptstyle{\sigma_{sum}=\sqrt{\sigma_\alpha^2+\sigma_\delta^2}}$, before and after DCR correction.}
	\label{fig2}
\end{figure*}

\begin{figure*}	
	\centering
\includegraphics[width=0.482\textwidth]{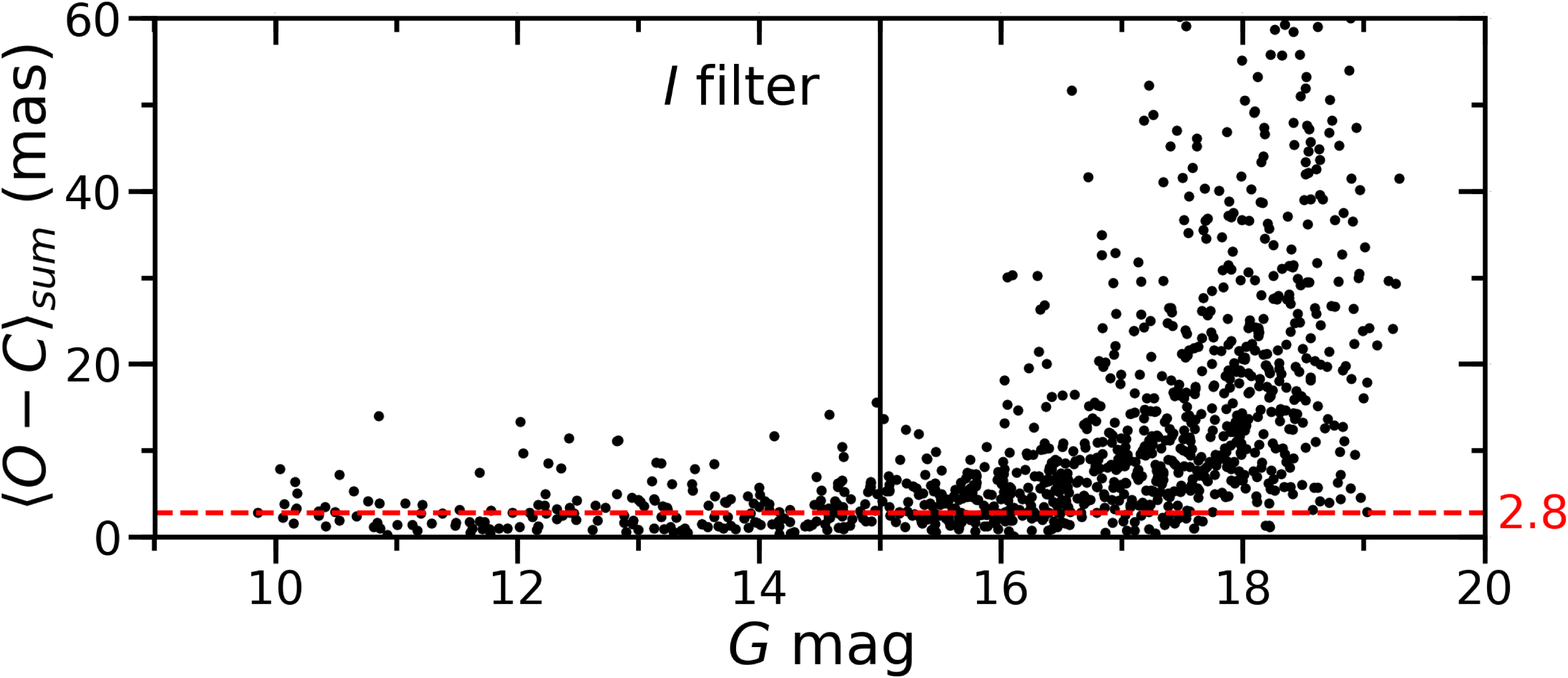}\quad
\includegraphics[width=0.482\textwidth]{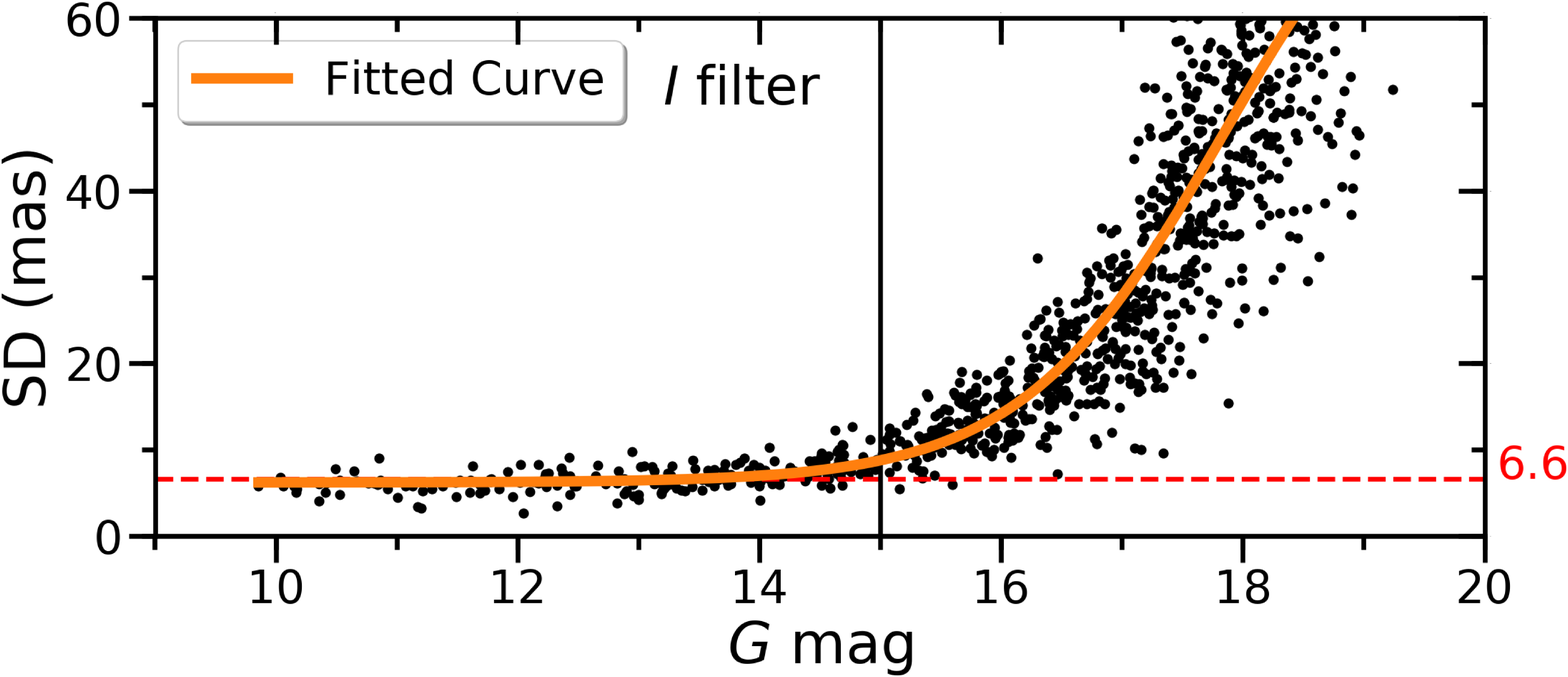}
		\caption{Reduction of observation set 5. The observations are taken under similar observational conditions (e.g. seeing) as observations set 1, but captured through the $I$ filter at small ZD. A few stars (5 in this figure) affected by stochastic factors (such as cosmic rays, binary stars, etc.) have been removed from the figure. A curve fitted by Equation~\ref{eq2} is shown in the right panel, and the detailed parameters of the curve are $A_1=0.0063, A_2=0.086, m_0= 17.86$ and $dm=0.83$.}
	\label{fig3}
\end{figure*}

Both the accuracy and precision of positional measurement are improved after DCR correction. Figure~\ref{fig2} shows the statistics of the $(O-C)$s before and after DCR correction for observation set 1, which are represented by red dots and black dots respectively. The standard deviation (SD) in the right panel of the figure is calculated by $\sigma_{sum}=\scriptstyle{\sqrt{\sigma_\alpha^2+\sigma_\delta^2}}$, and $\sigma_\alpha$ in the sense of the SD of $\Delta\alpha\cdot \cos\delta$. As shown in the figure, the improvement in precision is not very significant. This is because the influence of DCR on each positional measurement is very close when the ZD of an individual star changes little (see Table~\ref{tbl1}). In order to demonstrate that the effect of DCR has been eliminated, the reduction of observation set 5, which is captured through the $I$ filter at small ZD and atmospheric conditions (such as seeing) similar to observation set 1, is carried out here, and the results are shown in Figure~\ref{fig3}. The observations in $I$ band are less affected by DCR, and with being taken at a small ZD, the results shown in Figure~\ref{fig3} can be considered as unaffected by DCR \citep{kilic201211,dieterich2018dynamical,wang2018focus}. Comparing the null-filter results in Figure~\ref{fig2} with these $I$-filter results, we can see that the accuracy and precision of the null-filter observations after DCR correction show similar to those of $I$-filter observations, namely achieve the precision and accuracy not affected by DCR.

\begin{figure*}	
	\centering
\includegraphics[width=0.48\textwidth]{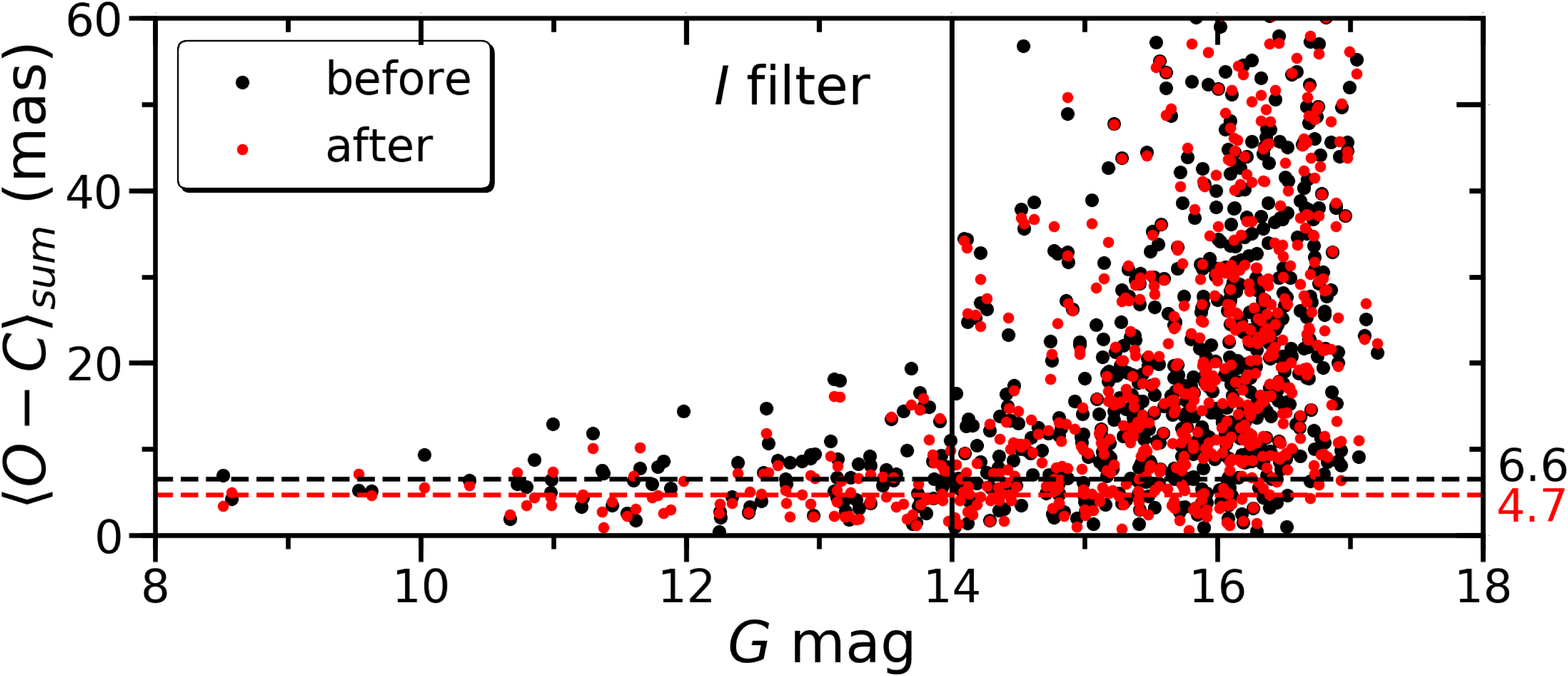}\quad
\includegraphics[width=0.482\textwidth]{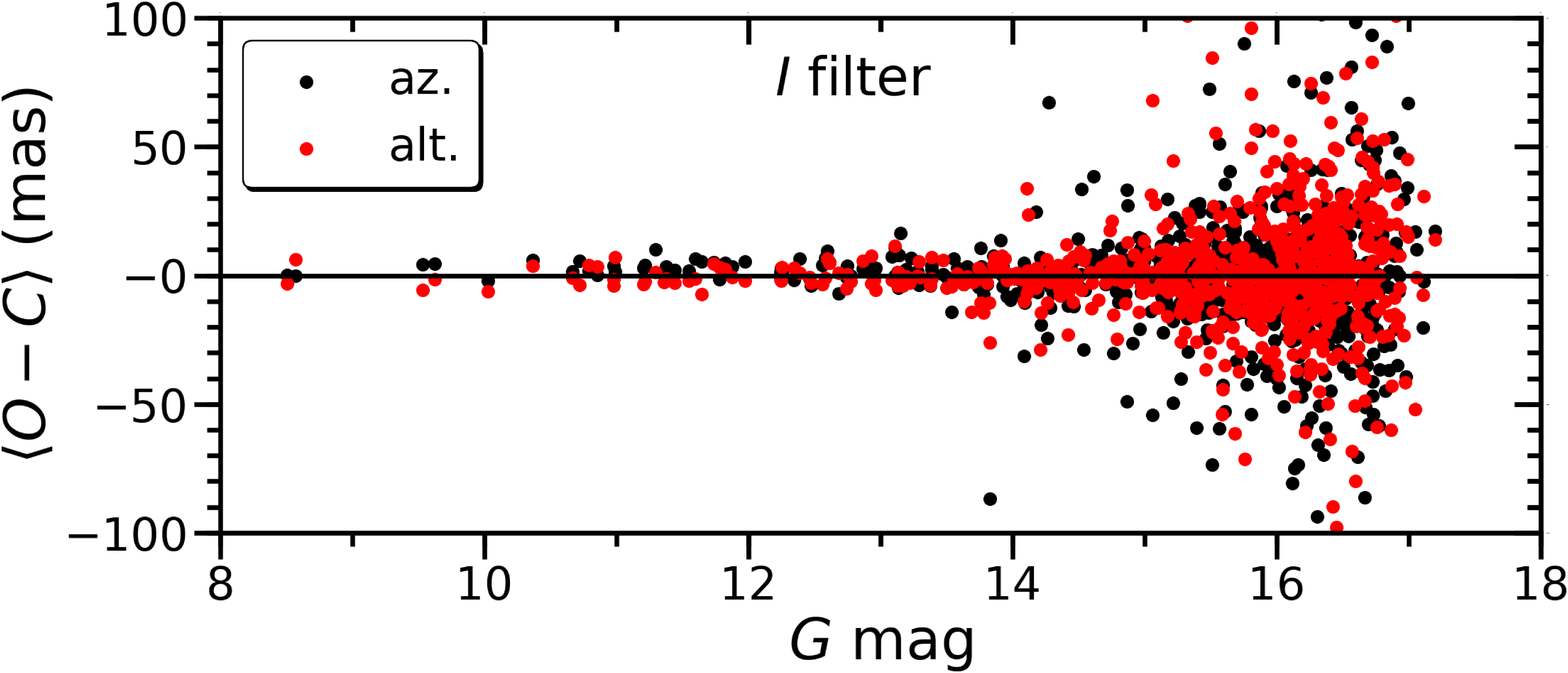}
\includegraphics[width=0.477\textwidth]{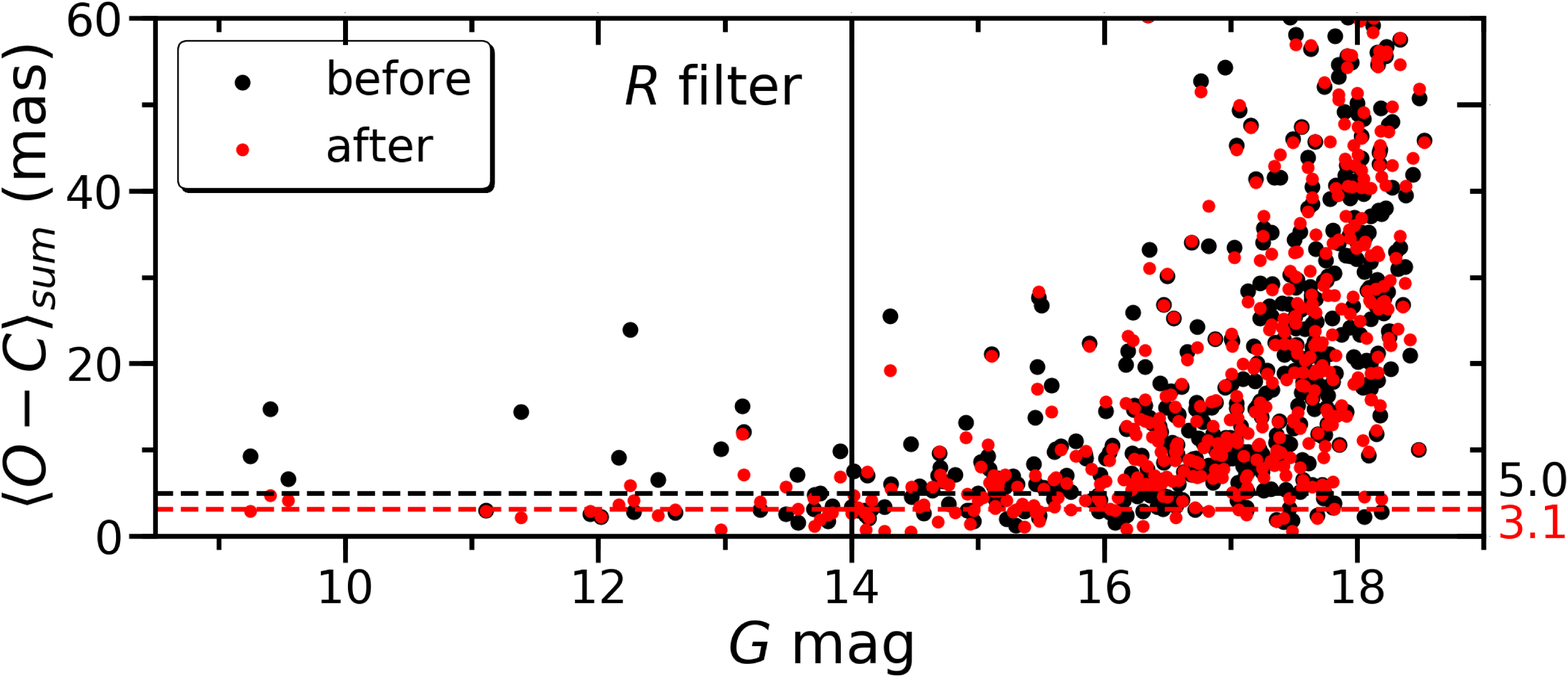}\;\;\;
\includegraphics[width=0.477\textwidth]{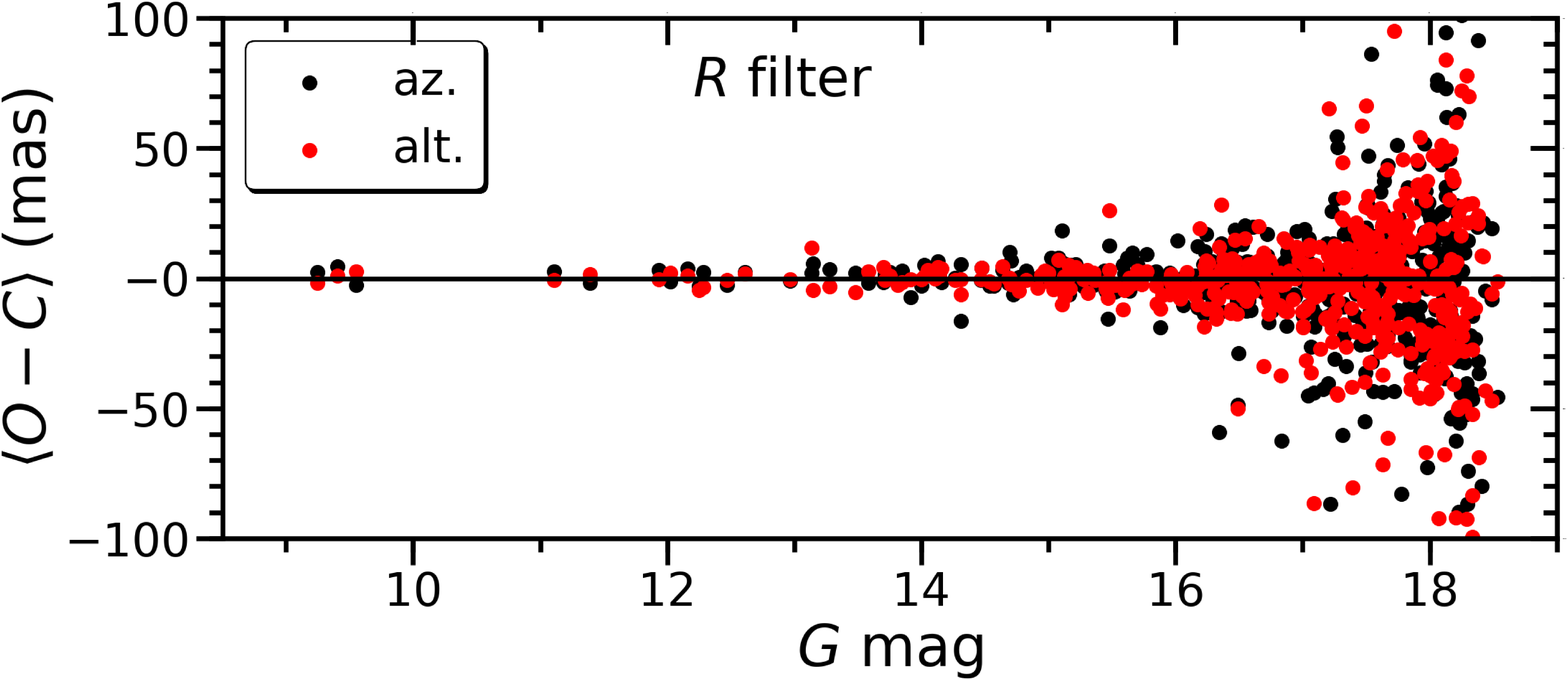}\\
\,\includegraphics[width=0.48\textwidth]{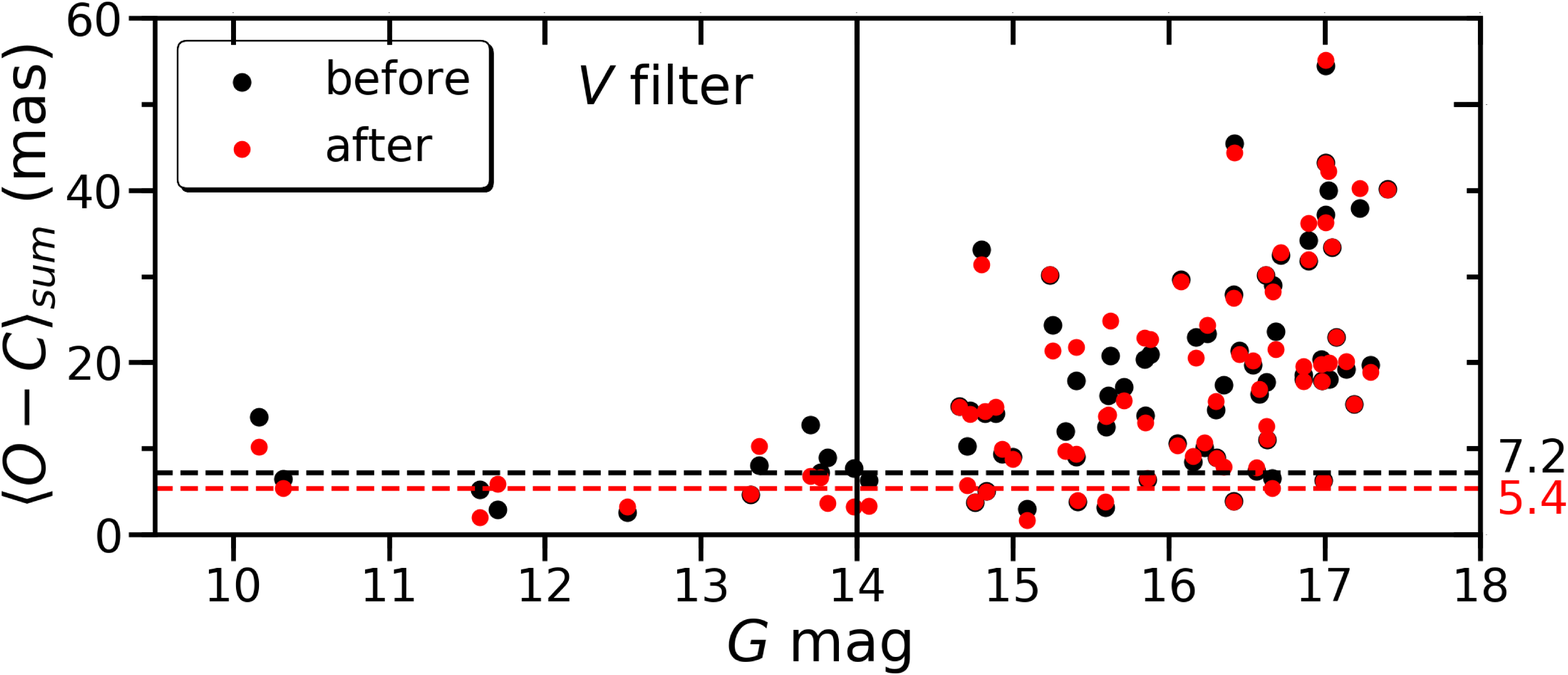}\quad
\includegraphics[width=0.485\textwidth]{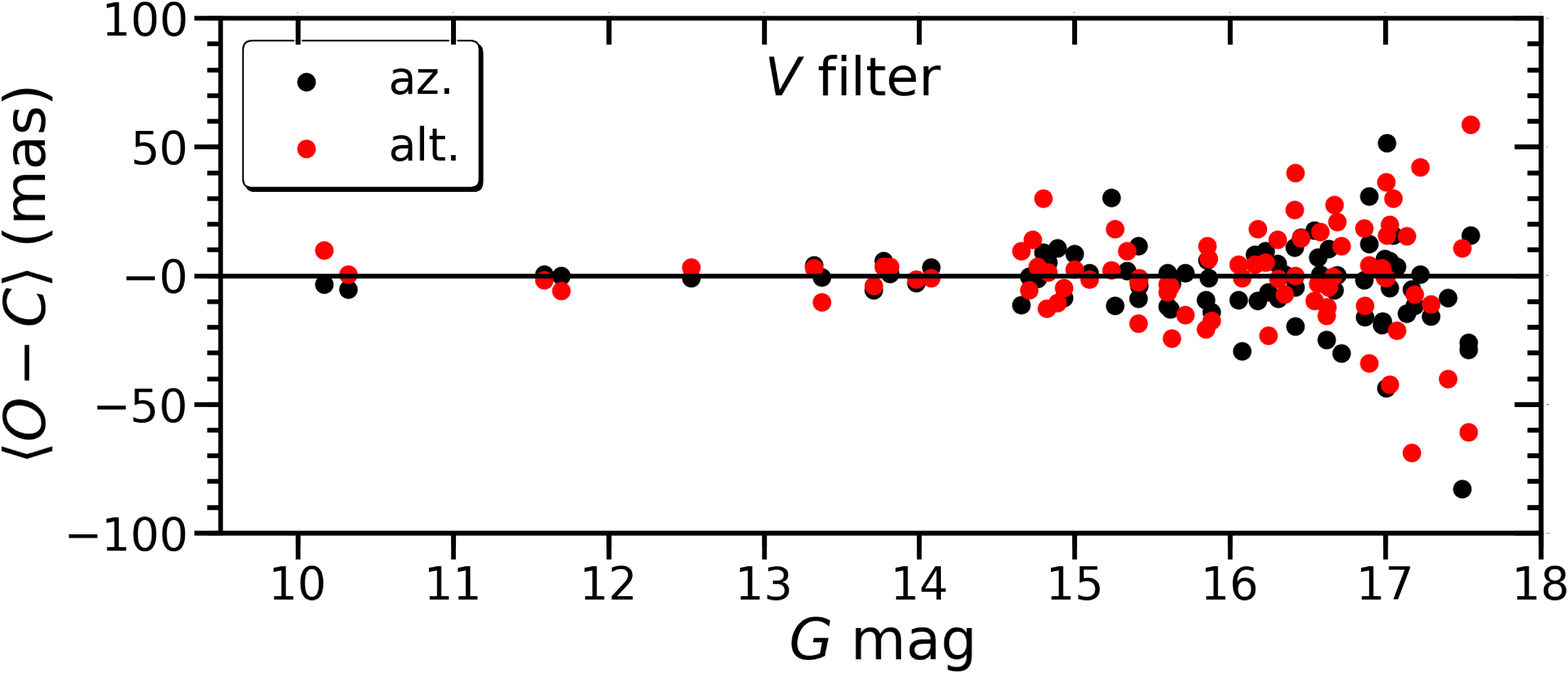}
\includegraphics[width=0.494\textwidth]{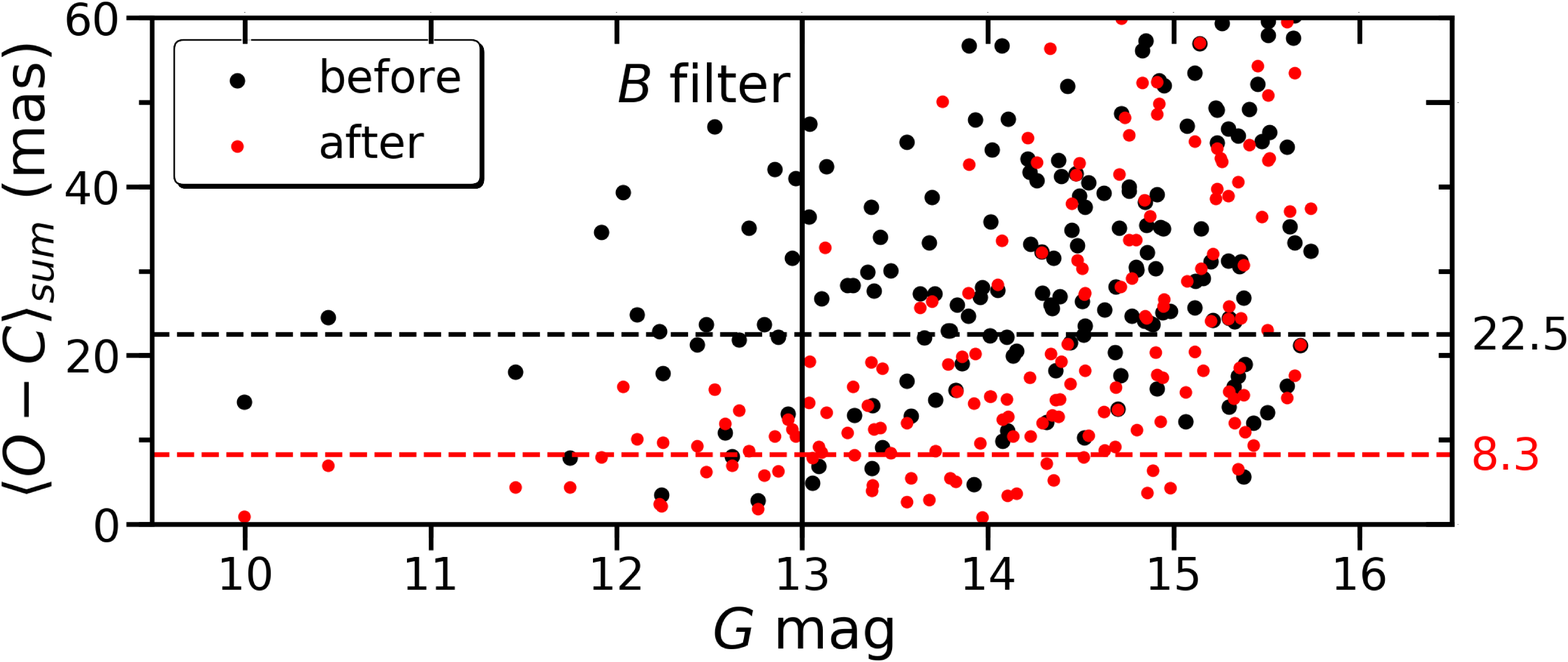}
\includegraphics[width=0.477\textwidth]{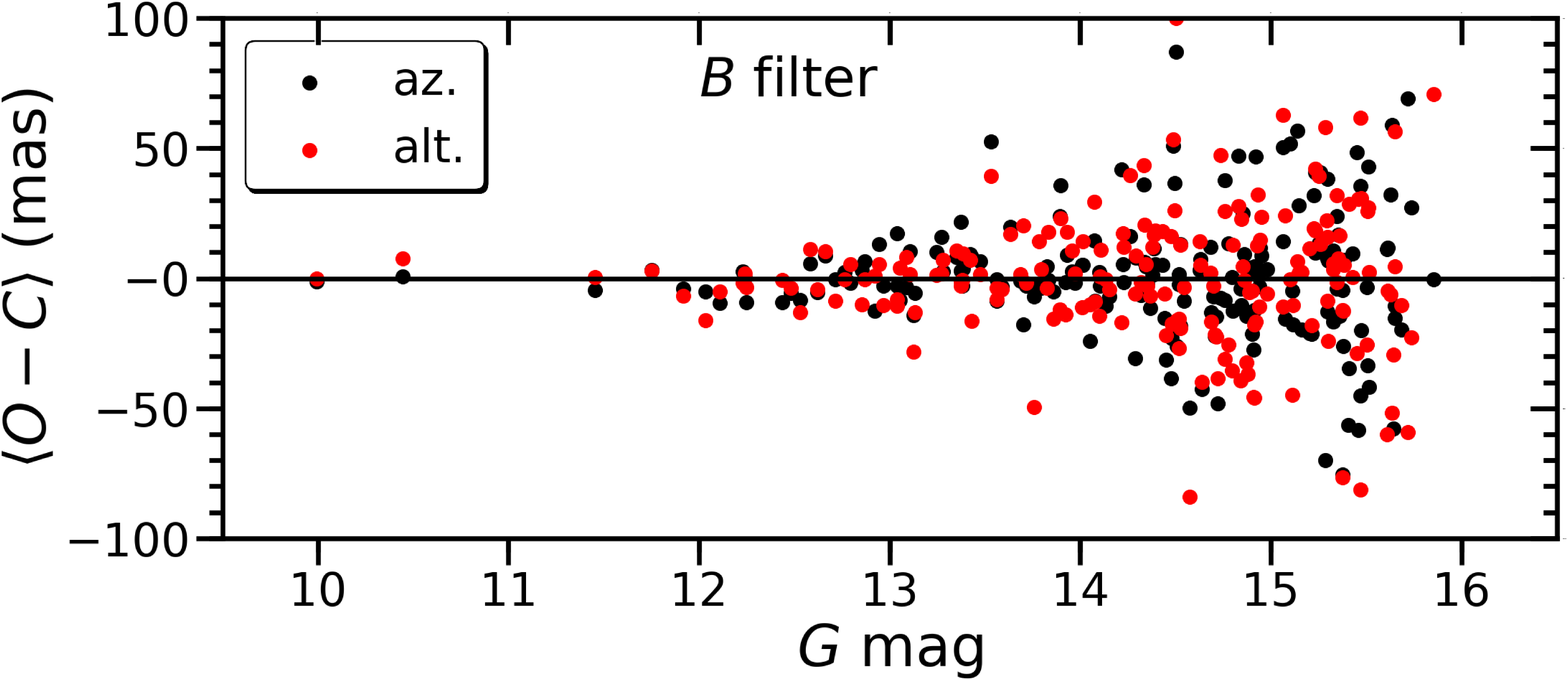}
		\caption{Left panels: comparison of positional bias before and after DCR correction for observations taken through different filters by the 1-m telescope at Yunnan Observatory (observation sets 6 to 9). The improvement of the V-filter observations is not obvious since they are captured at small ZD. Right panels: comparison of the dispersion of $\langle O-C\rangle$s in altitude and azimuth after DCR correction for the corresponding observation set in the left panel. } 	
	\label{fig4}
\end{figure*}

Observations taken through the Johnson $BV$ and Cousins $RI$ filters by the 1-m telescope at Yunnan Observatory are also reduced to verify the accuracy of DCR correction for these different filters. Specifically, the ability of our method to eliminate DCR effects is investigated by comparing the dispersion of $\langle O-C\rangle$s in altitude and azimuth after DCR correction. The results of the observation sets 6 to 9 are shown in Figure~\ref{fig4}, in which the left panels are the comparison of $\langle O-C\rangle_{sum}$ statistics before and after DCR correction, and the right panels the comparison between the $\langle O-C\rangle$s in altitude and azimuth after DCR correction. As can be seen from the figure, the measurement accuracy after DCR correction is improved for all the filters, and the dispersion of the $\langle O-C\rangle$ in altitude is the same as that in the unaffected azimuth direction.

 We found that the impact of atmospheric conditions to DCR would not be large for a given observatory. This is investigated by the DCR calibrations for observations over four consecutive nights captured through the $B$ filter on the 2.4-m telescope at Yunnan Observatory (observation sets 11 to 14 in Table~\ref{tbl2}), the DCR parameters of these observations are listed in the last four lines of Table~\ref{tbl3}. We can see from the results that the parameter related to the systematic error caused by DCR (i.e. $a_2$) is stable in four nights, with an average of 47.1 mas$\cdot mag^{-1}$. Although having a standard deviation of 3 mas$\cdot mag^{-1}$, it is not only affected by the changes in atmospheric conditions, but also by positional measurement errors of the calibration stars.
The precision of $a_2$ here is comparable to that of the slope parameters for individual stars in \citet{monet1992us}, which is devoted to the measurement of stellar parallax and so requires a stricter precision. That is to say, the lack of temperature and pressure information during the observation period will not cause problems in DCR calibration.
Furthermore, we found that the DCR of a particular filter would not change significantly for years (see lines 1 to 4 in Table~\ref{tbl3}, and the DCR solution of observation sets 8 and 10 taken through $V$ filter), even though other factors (such as seeing and pressure) may also be introduced to affect the stability of DCR solution.
However, since the filters in the same band may have different response curves, e.g. the $B$ filter on the 1-m and 2.4-m telescopes (see $a_2$ of observation sets 9 and 11 in Table~\ref{tbl3}), a recalibration of DCR for the selected filter would be better when DCR correction needs to solve precisely.

\begin{table*}
\begin{center}
	\caption{Detailed parameters of DCR equation derived from the observations in Table~\ref{tbl1}\label{tbl3}. The first four columns are the same as Table~\ref{tbl1}. Column (5) and (6) list the fitted parameters of Equation~\ref{eq5} and the standard deviation errors on these parameters.
}
\begin{tabular}{@{}c*{15}{c}}
\hline\hline
ID&Obs Date     &Filter&Telescope&$a_1$&$a_2$ \\
  &    (y-m-d) &  &&   (mas)&	(mas$\cdot$mag$^{-1}$) \\
 (1)&(2)&(3)&(4)&(5)&(6) \\
\hline
1  &  2011-02-26    & $N$     & 1-m   &  73$\pm$0.4& 	$-94.3\pm$0.4 \\
2  &  2012-10-14    & $N$     & 1-m   &  100$\pm$0.7&	$-92.2\pm$0.6   \\
3  &  2013-05-13    & $N$  & 1-m   &  117$\pm$1.9 &	 $-88.9\pm$1.4 \\
4  &  2014-10-17,18 & $N$    & 1-m    &  129$\pm$2.2 &	$-86.1\pm$1.4   \\
6  &  2012-04-18    & $I$     & 1-m   &  12$\pm$0.4&	$-7.1\pm$0.3  \\
7  &  2017-05-28,29 & $R$     & 1-m   &  30$\pm$0.9&	$-23.1\pm$0.6  \\
8  &  2015-12-21    & $V$     & 1-m   &  21$\pm$1.5&	$-18.5\pm$1.3  \\
9  &  2018-03-22    & $B$     & 1-m   &  48$\pm$1.5&	$-67.4\pm$1.5  \\
10 &  2019-03-30,31 & $V$     & 1-m   &  14$\pm$2.0&	$-17.7\pm$2.3  \\
11 &  2013-02-04    & $B$     & 2.4-m &  52$\pm$0.5&	$-50.9\pm$0.4  \\
12 &  2013-02-05    & $B$     & 2.4-m &  49$\pm$0.3&	$-48.7\pm$0.3  \\
13 &  2013-02-06    & $B$     & 2.4-m &  48$\pm$0.5&	$-44.7\pm$0.5  \\
14 &  2013-02-07    & $B$     & 2.4-m &  53$\pm$1.4&	$-44.1\pm$1.0  \\
\hline
\end{tabular}
\end{center}
\end{table*}

\subsection{Results of CTE correction}
\label{subsect:cteres}
\begin{figure*}	
	\centering
\includegraphics[width=0.49\textwidth]{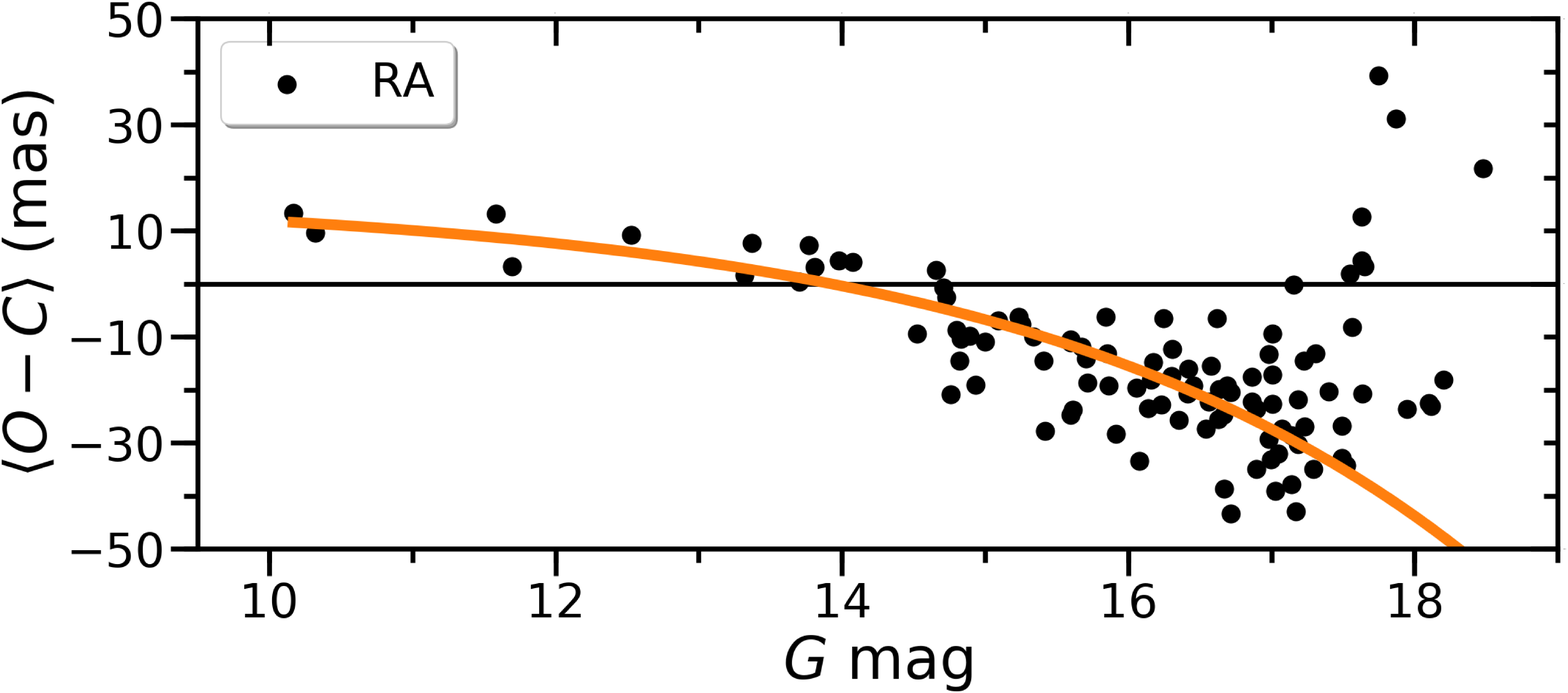}
\includegraphics[width=0.49\textwidth]{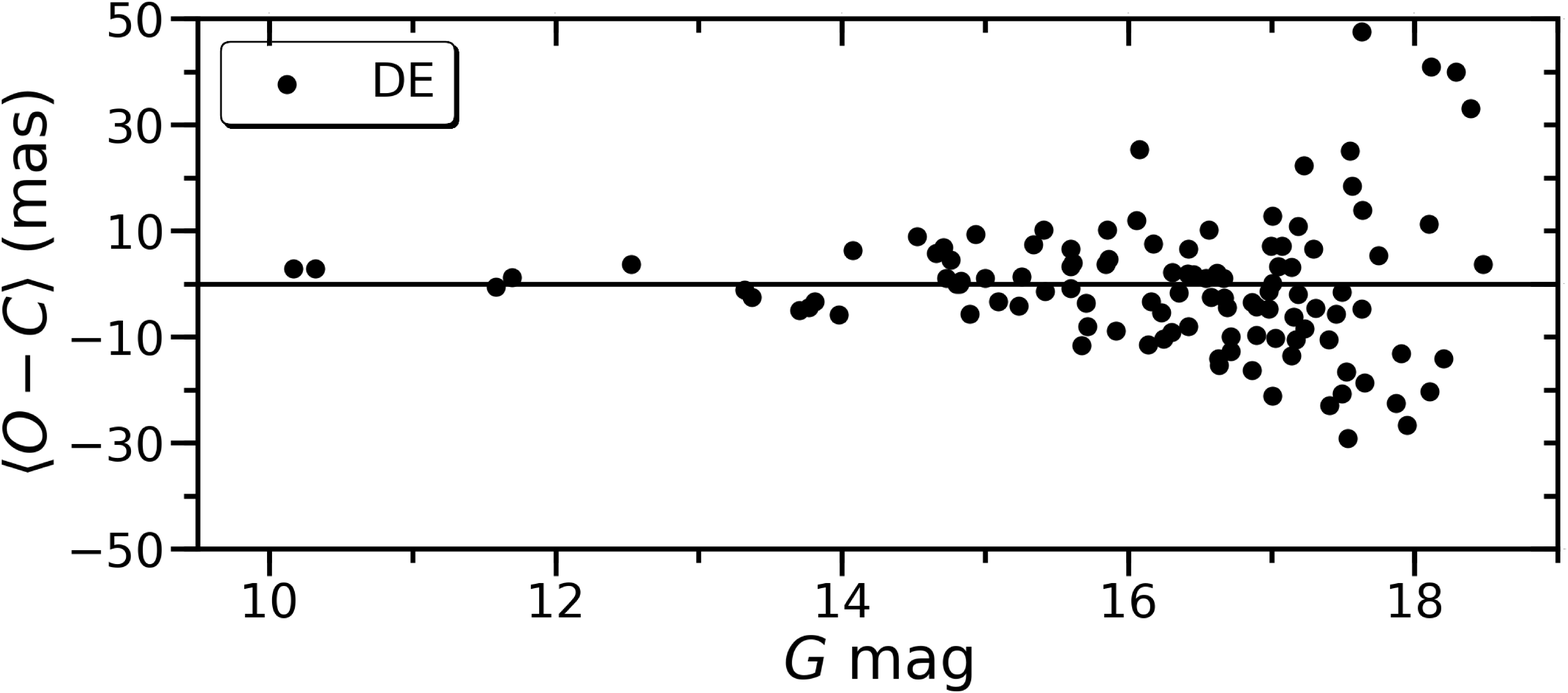}
		\caption{Reduction of observations affected by CTE issue (observation set 15). The left and right panels are the $\langle O-C\rangle$s in right ascension and declination respectively. The detailed parameters of the fitted curve in the left panel are $A_1=0.0167, A_2=-143.37, m_0= 42.61$ and $dm=3.17$.}
	\label{fig5}
\end{figure*}

\begin{figure*}	
	\centering
\includegraphics[width=0.49\textwidth]{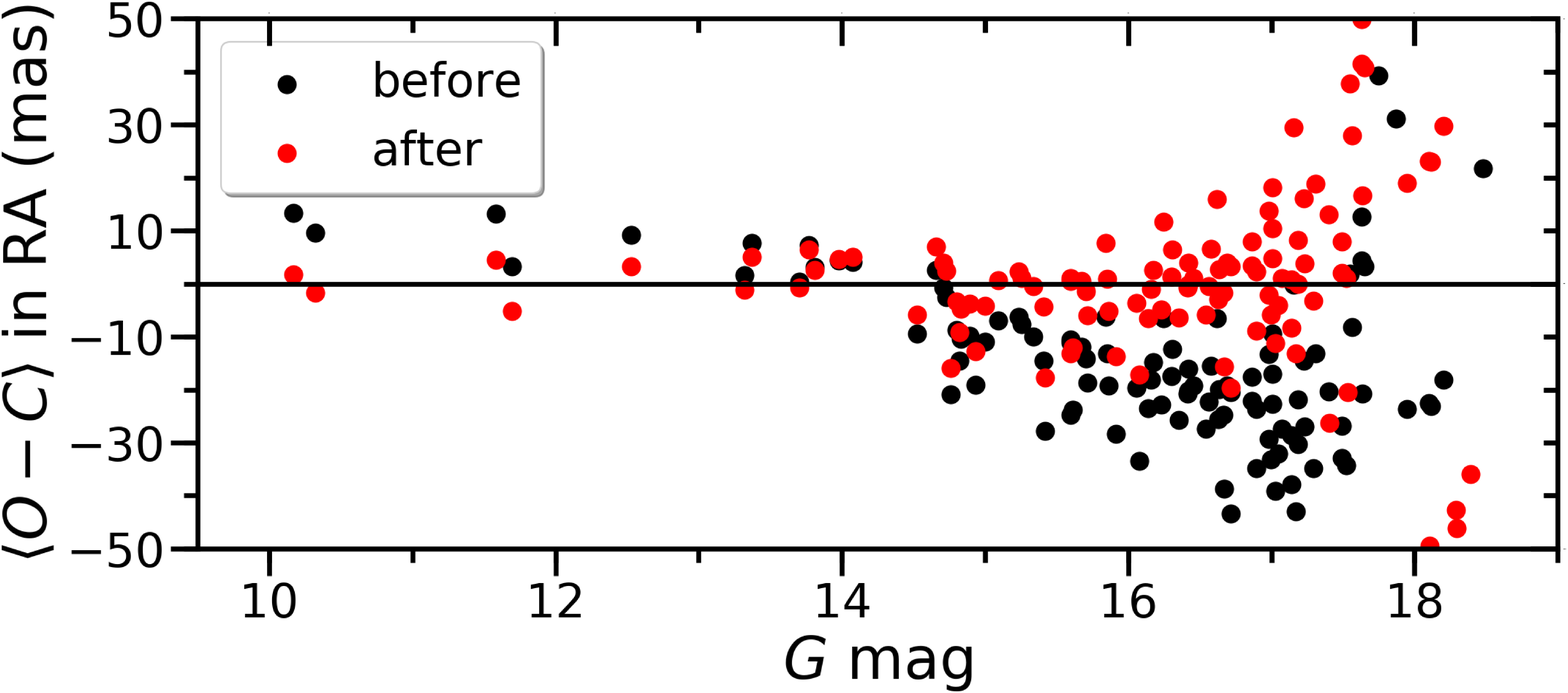}
\includegraphics[width=0.49\textwidth]{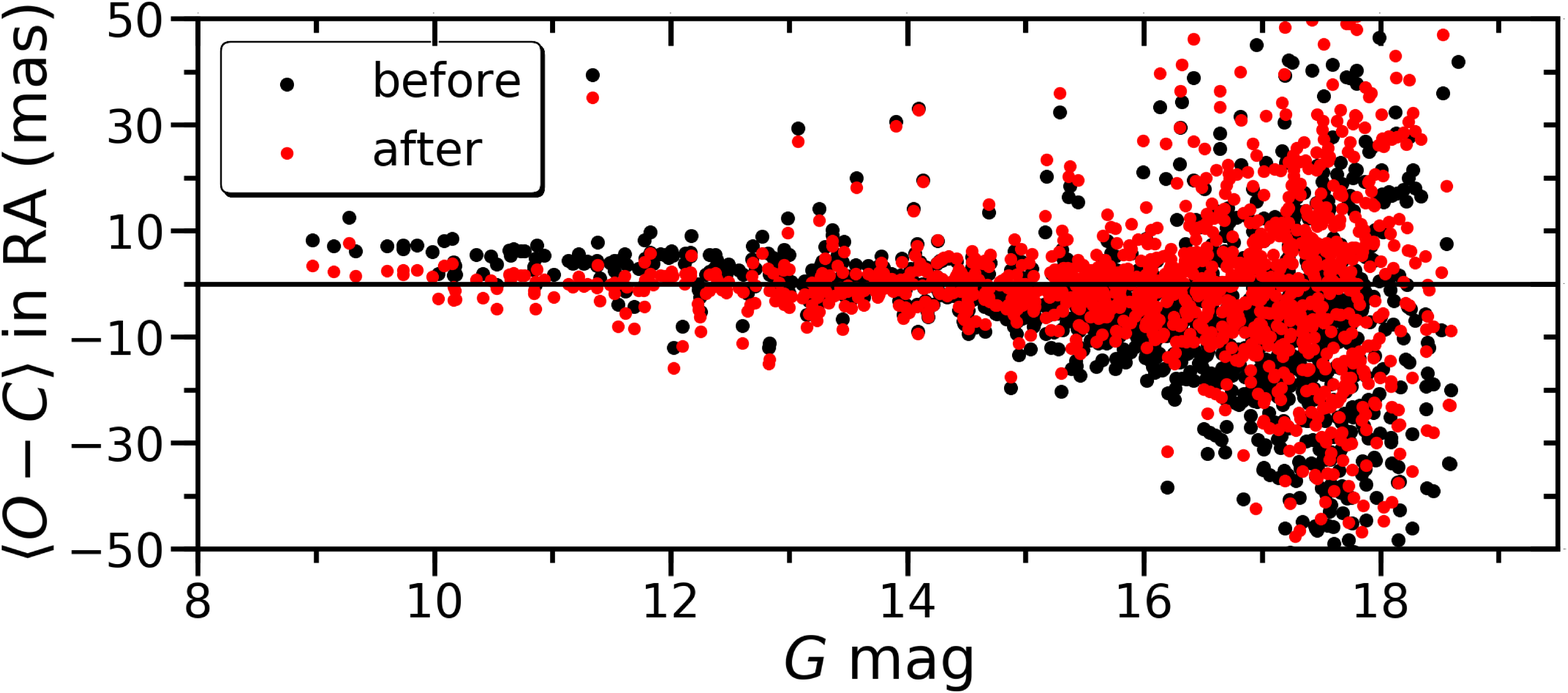}
		\caption{Comparison of the results before and after CTE correction for observation sets 15 (left panel) and 16 (right panel).} 	
	\label{fig6}
\end{figure*}

The systematic error caused by poor CTE is studied in this paper using observation sets 15 and 16 listed in Table~\ref{tbl1}. Figure~\ref{fig5} shows the reduction of observation set 15 that suffers from this error. From the figure, we can see that there is an obvious systematic trend in the right ascension (R.A.), which is the readout direction of the CCD. This trend is found to vary with time and should be tested on any observation set to avoid its impact to the results. Even so, it is not hard to handle this issue. Using the method mentioned in Section~\ref{sect:med}, the error can be well corrected. Specifically, Equation~\ref{eq2} is used to fit the $\langle O-C\rangle$s in R.A., and then the CTE corrected results can be obtained by subtracting the fitting value from the $\langle O-C\rangle$. 
For observation sets 15 and 16, the systematic errors before CTE correction can be up to about 30 mas and 16 mas, respectively. Figure~\ref{fig6} shows the $\langle O-C\rangle$s before and after CTE correction. From the trend of red dots in the figure, we can see that the error caused by poor CTE is eliminated after CTE correction.

\section{Conclusions}
In this paper, we proposed a method to calibrate differential color refraction (DCR) using the data of the astrometry and photometry (color $BP-RP$) given in Gaia DR2. This method is convenient since it only requires the observations taken through a single filter to calibrate the DCR of that filter. Moreover, even though few of the observations in this paper were specially taken for DCR calibration, all of them could be used to compute DCR with high precision by this method. Reduction of observations taken through the Johnson and Cousins' $BVRI$ filters and the null filter on the 1-m telescope at Yunnan Observatory shows that the accuracy of the data reduction after DCR correction achieves that not affected by DCR. The $B$-filter observations captured by the 2.4-m telescope at Yunnan Observatory are also processed. The results show that the DCR solution is stable in four consecutive nights, although the changes in atmospheric conditions (pressure and temperature) are ignored. The precision of parameter $a_2$ in four nights is better than 3 mas$\cdot$mag$^{-1}$, which means that the residual effect of DCR is less than 3 mas except for a few stars. In other words, this small residual can be realized under the condition of color index $|BP-RP|< 1$ and altitude $<45^\circ$. A long-term monitoring (over 4 years) of DCR for the null filter on the 1-m telescope led to a similar conclusion. Nevertheless, using our method to realize a recalibration of DCR for some selected filter would be more preferable when a better DCR solution is required.

The charge transfer efficiency (CTE) issue existed in the observations of the 1-m telescope at Yunnan Observatory is also investigated in this paper.
Two methods, namely the method of including magnitude terms in the reductions and the method used in the reduction of our observations (see Section~\ref{subsect:cteres}), can be used to handle the systematic error caused by poor CTE, and the better method can be selected from them according to the number of reference stars. Applying the correction for CTE to the reduction of observations, a systematic error up to 30 mas has been eliminated.

\section*{Acknowledgements}

This work was supported by the National Natural Science Foundation of China (Grant Nos. 11873026, 11703008, 11273014), by the Joint Research Fund in Astronomy (Grant No. U1431227) under cooperative agreement between the National Natural Science Foundation of China (NSFC) and Chinese Academy Sciences (CAS), and partly by the Fundamental Research Funds for the Central Universities. The authors would like to thank the chief scientist Qian S. B. of the 1-m telescope and his working group for their kindly support and help. And thank them for sharing the observations of AH Aur. This work has made use of data from the European Space Agency (ESA) mission \emph{Gaia} (\url{https://www.cosmos.esa.int/gaia}), processed by the \emph{Gaia} Data Processing and Analysis Consortium (DPAC, \url{https://www.cosmos.esa.int/web/gaia/dpac/consortium}). Funding for the DPAC has been provided by national institutions, in particular the institutions participating in the \emph{Gaia} Multilateral Agreement.

\section*{Data Availability}

The data underlying this article will be shared on reasonable request to the corresponding author.




\bibliographystyle{mnras}
\bibliography{DCRAndCTE} 


\bsp	
\label{lastpage}
\end{document}